%% file: Main.tex
\documentclass[conference]{IEEEtran}
\usepackage{caption}
\usepackage{algorithm}
\usepackage{algpseudocode}
\usepackage{graphicx}
\usepackage{subcaption}
\usepackage{bm}
\usepackage{mathtools}
\usepackage{amssymb}
\usepackage{multirow}

\usepackage{enumitem}

\setenumerate[1]{itemsep=0.5pt,partopsep=0pt,parsep=\parskip, topsep=2pt, leftmargin=12pt,}

\begin{document}

\title{\huge{BDFA: A Blind Data  Adversarial Bit-flip Attack on Deep Neural Networks}
}

\author{\IEEEauthorblockN{Behnam Ghavami, Mani Sadati, Mohammad Shahidzadeh, Zhenman Fang, Lesley Shannon}
 \IEEEauthorblockA{Simon~Fraser~University, Burnaby, BC, Canada\\
 Emails: \{behnam\_ghavami, zhenman, lesley\_shannon\}@sfu.ca}}

\maketitle
\input{abstract}
\input{intro}
\input{relate}
\input{proposal}

\input{results}
\input{concl}

\bibliographystyle{IEEEtran}
\bibliography{references}	

\end{document}

%% file: abstract.tex
\begin{abstract}
Adversarial bit-flip attack (BFA) on Neural Network weights can result in catastrophic accuracy degradation by flipping a very small number of bits. A major drawback of prior bit flip attack techniques is their reliance on test data. This is frequently not possible for applications that contain sensitive or proprietary data.
In this paper, we propose Blind Data Adversarial Bit-flip Attack (BDFA), a novel technique to enable BFA without any access to the training or testing data.
This is achieved by optimizing for a synthetic dataset, which is engineered to match the statistics of batch normalization across different layers of the network and the targeted label. Experimental results show that BDFA could decrease the accuracy of ResNet50 significantly from 75.96\% to 13.94\% with only 4 bits flips.

\end{abstract}


%% file: intro.tex
\section{Introduction}\label{sec:intro}

In recent years, Deep Neural Networks (DNNs) have achieved tremendous results on different computer vision and speech recognition tasks such as image classification, object detection, and segmentation.
As DNNs become more popular and applicable in real-world scenarios, their security and safety issues are also becoming crucial. Therefore, it is very important to study the vulnerability and safeguard of such DNN-based systems under various attacks. 

Several significant DNN security issues have recently been raised in connection with various applications and components. The most widely researched security concern of DNN stems from its malicious input, also known as an adversarial example \cite{szegedy2013intriguing,goodfellow2014explaining,moosavi2016deepfool,carlini2017towards,moosavi2017universal,hayes2018learning,chaubey2020universal} which fool DNNs to misclassify crafted inputs with imperceptible perturbations. 

Recently, a new class of attacks has raised further security concerns on DNNs known as adversarial parameter attacks via the development of fault injection attack on the storage of DNN parameters \cite{liu2017fault, breier2018practical, rakin2019bit, yao2020deephammer,  zhao2019fault, rakin2020t}. This type of attacks try to perturb DNN's parameters in the memory via bit flips and cause the DNN to malfunction. These malicious bit flips have been realized in DNN accelerators via well know  RowHammer attack on the DRAM containing the model parameters \cite{yao2020deephammer}. RowHammer attack has been shown to maliciously flip the memory bits in DRAM in a software manner without being granted any data write privileges.

Note that adversarial parameter perturbations is an active area of research in security analysis of DNN models \cite{liu2017fault, breier2018practical, rakin2019bit, yao2020deephammer,  zhao2019fault, rakin2020t}. However, practical applying adversarial perturbations on network is not trivial in real applications, as existing works on bit flip attacks mainly focus on the white-box setting, where the adversary is capable to access all the information about the target model and data. Under white-box assumption, an adversary has access to the network architecture, weight values and one batch of test data. While network architecture and parameters information can be potentially extract 
by leveraging side-channel model-extraction  techniques\cite{breier2021sniff}\cite{chabanne2021side}\cite{batina2021sca,hua2018reverse,oh2019towards,oh2019towards,gongye2020reverse,batina2019csi,wei2020leaky,jha2020deeppeep,gongye2020reverse,zhu2021hermes,rakin2021deepsteal}, however, still the main challenge to applying prior adversarial bit flip attacks (BFAs) is that, the perturbing entity, i.e., the adversary, should be able to access on network test/validation data. This is because in BFA, as will be introduced, the adversary should determine the vulnerable bits and hence, needs to rank the sensitivity of every attackable bit over all the DNN's parameters. Training/Test data as are often unavailable in many scenarios due to privacy and security issues, such as medical and confidential scenarios. An example application case is health care information which can not be uploaded to the cloud due to various privacy issues and/or regulatory constraints. We refer to this case as the blind-data setting.

In this paper, we are the first to introduce an effective blind-data bit attack that can attack the DNN parameters and flip a few of them, causing the DNN to malfunction completely. We generate synthetic data similar to the training dataset using the knowledge of deep learning datasets in general and the DNN architecture. We obtain the Distilled Data by solely analyzing the trained model itself. This work is a step forward towards the fully black-box parameter attack. Experimental results show that the Blind Data Bit Flip attack (BDFA) can perform similar results to data-dependent Bit-Flip Attack (BFA) \cite{rakin2019bit}. For example, on ResNet50 architecture  trained on the CIFAR100 dataset, BDFA can crush the DNN model down to the accuracy of 11\% by just flipping 8 bits in memory.

The rest of this paper is as follow: Section 2 gives an overview on prior DNN bit flip attacks. Section 3 contains the proposed blind data bit flip attack. Section 4 presents the experimental results. Finally, Section 5 concludes the paper.

%% file: relate.tex
\section{Related Work}
\label{sec:related}

Memory fault injection attacks have been used to perturb the major parameters of a DNN such as weights and biases mainly focus on reducing the overall prediction accuracy to be as low as random guess. The first work that explored memory fault injection of a DNN hardware to achieve misclassification was done by Liu et al. \cite{liu2017fault}. They provided attacks on a certain bias term of a neuron.
Breier et al. \cite{breier2018practical} experimentally showed what types of memory fault attacks are achievable in practice. They injected faults into the activation function of the DNN to missclassify a target input. 
Rakin et al. \cite{rakin2019bit} presented a method to find the specific memory fault patterns that can cause important destruction to the DNN accuracy.
Taking advantage of the well-known row hammer attack \cite{kim2014flipping}, Yao et al. \cite{yao2020deephammer} attempted to attack a DNN hardware where the network weights are stored in DRAM.
Lately, Zhao et al. \cite{zhao2019fault} introduced a bit flipping attack on a DNN classifier in order to stealthily misclassify a few predefined inputs. 
Ghavami et al. \cite{ghavami2021stealthy} presented a new type of stealthy attack on DNNs to circumvent the algorithmic defenses: via smart bit flipping in DNN weights, they reserve the classification accuracy for clean inputs but misclassify crafted inputs even with algorithmic countermeasures. 
Rakin et al. \cite{rakin2020t} also introduced an adversarial bit flip attack on DNN models whose main goal is to identify the weights that are highly associated with the misclassification of a targeted output.


\textbf{Limitations of previous works:} All prior bit flip attacks require access to the original testing dataset for finding most vulnerable bits where such requirement may not be applicable in all scenarios.

 
 

%% file: proposal.tex
\section{Proposed Attack}
\label{sec:proposed}
In this section, we present the Blind Data Bit-Flip Attack (BDFA) to maliciously cause a DNN system malfunction through flipping extremely small amount of most vulnerable bits of weights. To identify the most vulnerable bits, there is a need to compute the gradient of each bit with respect to the DNN's loss function and some input data. Since we assume that the training/testing dataset is not accessible, we generate a batch of synthetic test data using the trained network architecture itself. Our main idea is inspired by prior work in data free DNN quntization \cite{cai2020zeroq}.

\subsection{Bit-Flip Attack \cite{rakin2019bit}}
Bit-flip attack (BFA) \cite{rakin2019bit} uses the gradient ranking and a progressive bit search to find the most vulnerable bits of a network.
\subsubsection{Problem Formulation}
We denote $W$ as a vector of length $n$ containing all the q-bit quantized and attackable parameters in DNN and $B$ as the vector of binary values, representing all of the bits in $W$ \cite{rakin2019bit}.
BFA tries to find a set of parameters $B'$, which has the closest hamming distance $D$ to $B$ and causes the network to malfunction. In other words, it tries to maximize the loss between real parameters $B$ and perturbed parameters $B'$ where the distance between them is smaller than a constant $C$ which is the max number of bit flips that can be performed \cite{rakin2019bit}:
\begin{equation}
    \begin{split}
    \max_{B'}\; \mathcal{L}(&H(B';X),Y) - \mathcal{L}(H(B;X),Y) \\
    & s.t.\;\;\; D(B',B) < C
    \end{split}
\end{equation}
where $X$ is the input batch and $Y$ is the target output of that batch and $L$ is loss function on $D$ through adjusting DNN parameters.

\subsubsection{Vulnerable bit finding}
Using a batch of test data, BAF finds the most vulnerable bits. It first ranks the bits by their gradient, and in the next step tries to find the most vulnerable bits.
It ranks all the network's bits $B$ by absolute value of their gradient with respect to the loss $L$. For this purpose, it first computes the loss function $L$ by providing input data X and output targets $Y$ computed in the section B. Then, it calculates the gradients with respect to the loss through back propagation \cite{rakin2019bit}:
\begin{equation}
    \mathcal{L} = \frac{1}{N} \sum^{N}_{i=1} \;\;\mathclap{f}\;( H(X_{i};B) , Y_i)
\end{equation}
\begin{equation}
    \overrightarrow{\nabla}\mathcal{L} = \{\frac{\partial \mathcal{L}}{\partial b_{0}},...,\frac{\partial \mathcal{L}}{\partial b_{N}}\}
\end{equation}
For finding the most vulnerable bits BFA uses progress bit search that has two main steps. a) Inner-layer search: In this step, it finds the most vulnerable bits with the use of gradient ranking in every layer and compute the model loss after flipping them. B) Cross-layer search: In this step, it chooses the layer that increased the loss more than other layers and flip the selected bit in that layer. It performs this process several times until it causes the network to malfunction with a small number of bit flips.

\subsection{Attack Oriented Generating Synthetic Test Data}
As shown in the previous section, BFA needs a batch of input data and the corresponding labels in the form of $\{(x_{1},y_{1}),(x_{2},y_{2}),...,(x_{n},y_{n})\}$ for computing the loss function which is not accessible in all scenarios. To address this issue, a very naive approach would be to generate random input data from a Gaussian distribution and feed it into the model. This method, however, is incapable of capturing the correct statistics of the training/testing data used for computing the gradients.

In order to generate proper input data to perform attack, we use multiple similarity measures to reconstruct statistically similar data samples to dataset $D$ by starting from randomly generated samples.

\subsubsection{Input similarity}
Starting from random data, in order to make them statistically similar to training data, they should have a close mean and variance to the data samples in $D$. We set the mean and variance of the initial random data to 0 and 1, respectively. This is because almost all of the deep learning systems use a normalized input to get a more standard and accurate model.

\subsubsection{Batch normalization layer statistics}
Another statistical information comes from the batch normalization layers. Each batch normalization layer contains statistical channel-wise information (mean and variance) of its input neurons during the training process. As a result, by making the statistics of hidden neurons close to the pre-stored statistics of training data, we can have more similarity between the generated data and the training samples. The formulated batch-norm similarity can be shown as \cite{cai2020zeroq}: 
\begin{equation}
    \tilde{\mu}_l(c) = \mathbf{E}_x [(\frac{1}{h * w} . featuremap^c_{l} (x)]
\end{equation}
\begin{equation}
    \tilde{\sigma}^2_l(c) = \mathbf{E}_x [(\frac{1}{h * w} . (featuremap^c_{l} (x) -  \tilde{\mu}(c))^2]
\end{equation}
where $featuremap^c_{l} (x)$ represents the $cth$ input channel in the $lth$ bach normalization layer given input data $x$. Note that both $\tilde{\mu}_l$  and $\tilde{\sigma}^2_l$ are vectors with length $C_l$ which is the number of input channels in $lth$ batch normalization layer. We calculate $\tilde{\sigma}_l$, the standard deviation of each channel by taking the square root of the elements in the vector $\tilde{\sigma}^2_l$.

The generated batch should have a close $\mu$ and $\sigma$ to the $\tilde{\mu}$ and $\tilde{\sigma}$ that were computed in the training process. In order to estimate the similarity of generated data and the training data, we use the mean squared error as the loss function. By minimizing this loss function, we decrease the Euclidean distance of statistical information in batch-norm layers between the training data and the synthetic data \cite{cai2020zeroq}:
\begin{equation}
    \min_{X}\; \mathcal{L}oss_{BN}(X) = \sum_{i=0}^{L} ||\tilde{\mu}_i - \mu_i||^2_2 + ||\tilde{\sigma}_i - \sigma_i||^2_2
\end{equation}

\subsubsection{Label similarity}
To get the gradient of each bit in every parameter, we need to compute the loss function. Based on Equation 2, in order to compute the loss function, each input data should have a ground-truth label. So, we need to assign a label to each generated data sample. Also, every parameter in the network is trained to minimize the loss function over training data. Therefore, to have similar distilled data and training data, we train distilled data in a way that the model's loss function is minimized with respect to the given input and the ground-truth label. However, instead of adjusting DNN parameters, we adjust the input data to minimize the loss function. Since the generated data are random at the beginning and do not have any labels, we can randomly assign labels to each input and train them to minimize the model loss.
\begin{equation}
    \min_{X^r} \;\mathcal{L}oss_{DNN}(X) = \frac{1}{N} \sum^{N}_{i=1} \;\;\mathclap{f}\;( Y^{'}_{i} , Y_i)
\end{equation}

\subsubsection{Combining together}
We define the distillation loss as a combination of $ \mathcal{L}oss_{DNN}$ and $ \mathcal{L}oss_{BN}$, and try to change the random training batch such that every input has the $mean=0$ and $var=1$:
\begin{equation}
    \min_{X^r} \;\mathcal{L}(X) = \alpha \;.\; \mathcal{L}oss_{BN}(X) + \;\beta \;.\; \mathcal{L} oss_{DNN}(X)
\end{equation}
$\alpha$ and $\beta$ are hyper parameters for the loss function to balance the effect of each part of the loss function. 


The pseudo-code provided for the task is presented in Algorithm \ref{Alg:main}. In this algorithm, given model M and knowing that the data shape is $N*C*H*W$, we want to generate a batch of data X (line 2). Note that $N$ is the batch size, and each input data has a shape of $C*H*W$. The algorithm begins by generating a random batch of data from the normal distribution with $\mu = 0$ and $\sigma = 1$. Line 3 and 4 store computed mean and standard deviation of each BN layer, which were calculated and saved in those layers, during the training process. In line 5, as mentioned in section C.3, we randomly assign each data in the batch to a ground-truth label, and we will use these labels to train our distilled data according to equation 6. Line 6 to 12 is the main loop for the generation of data. Like every other deep learning training process, we start each iteration with a forward propagation and compute $\mu$ and $\sigma$ and the outputs of DNN, using our data X (line 7). In lines 8 and 9, we compute the two parts of the final loss function according to equations 5 and 6. Then we calculate the total loss by combining these two parts and adding the hyper-parameters  to balance the loss function. Finally, in line 11, we do the back-propagation and update data X. By doing this for enough iterations (e.g., 500), we can produce our distilled data and use it to attack the DNN model.

\begin{algorithm}
\caption{Synthetic Data Generation}\label{alg:euclid}
\hspace*{\algorithmicindent} \textbf{Input:} A Deep learning model M with L layers of BN\\
\hspace*{\algorithmicindent} \hspace{10mm}  shape=(N=batch size,C=3,H=32,W=32)\\
\hspace*{\algorithmicindent} \textbf{Output:}Generated data X 
\begin{algorithmic}[1]
\Procedure{Generate Data}{$M$,$Shape$}
    \State $\overrightarrow{X}$ $\gets$ RandomNormalizedData(Shape)
    \State $\tilde{\mu}_{\forall{j}\in{1,2,..,L}}$ $\gets$ computed mean in the training process
    \State $\tilde{\sigma}_{\forall{j}\in{1,2,..,L}}$ $\gets$ computed std in the training process
    \State $y$ $\gets$ GenerateRandomLabel(N)
    \For {$iteration=1,2,\dots$}
        \State ${\mu}_j$,${\sigma}_j$,$y^{'}$ $\gets$ ForwardProp($\overrightarrow{X}$,M)\Comment{$\forall{j}\in{1,2,..,L}$}
        \State $loss_{BN}$ $\gets$ BNLOSS(${\mu}$,${\sigma}$,$\tilde{\mu}$,$\tilde{\sigma}$)\Comment{equation 5}
        \State $loss_{DNN}$ $\gets$ DNNLOSS($y$,$y^{'}$)\Comment{equation 6}
        \State loss $\gets$ $\alpha.loss_{BN} + \beta.loss_{DNN}$
        \State $\overrightarrow{X}$ $\gets$ updated $\overrightarrow{X}$ by BackProp(loss)
    \EndFor\label{euclidendwhile}
    \State \textbf{return} $(\overrightarrow{X},y)$\Comment{$\overrightarrow{X}$ is the generated artificial input batch and y is the target output}
\EndProcedure
\end{algorithmic}
\label{Alg:main}
\end{algorithm}

%% file: results.tex
\section{Experimental Results}
\label{sec:results}
\subsection{Experimental Setup}
\subsubsection{Datasets}
We used CIFAR-10, and CIFAR-100 \cite{CIFAR-10}, popular datasets for image classification. We use these datasets to train our models. Both CIFAR-10 and CIFAR-100 contain 60000 RGB images with a size of $32*32$.

\subsubsection{DNN Architectures}
We chose VGG16 and ResNet50, which are two of the conventional CNN architectures. Both of these architectures use batch normalization layers to achieve better performance. VGG16 and ResNet50 have, respectively, 13 and 53 batch normalization layers. We implement these architectures in the Pytorch framework \cite{paszke2019pytorch} and use 8-bit quantization \cite{krishnamoorthi2018quantizing} for network parameters.

\subsubsection{Attack Assumptions}
In our experiments, we assume that we have full access to the network's parameters and architecture. However, contrary to previous papers, we do not assume having a batch of input data; Instead, we use generated distilled data as the inputs for DNN.

\begin{figure*}[!ht]
\vspace{-0.2in}
\centering
\captionsetup[subfloat]{position=top,labelformat=empty}
\subfloat[][]{
  \includegraphics[width=0.8\linewidth]{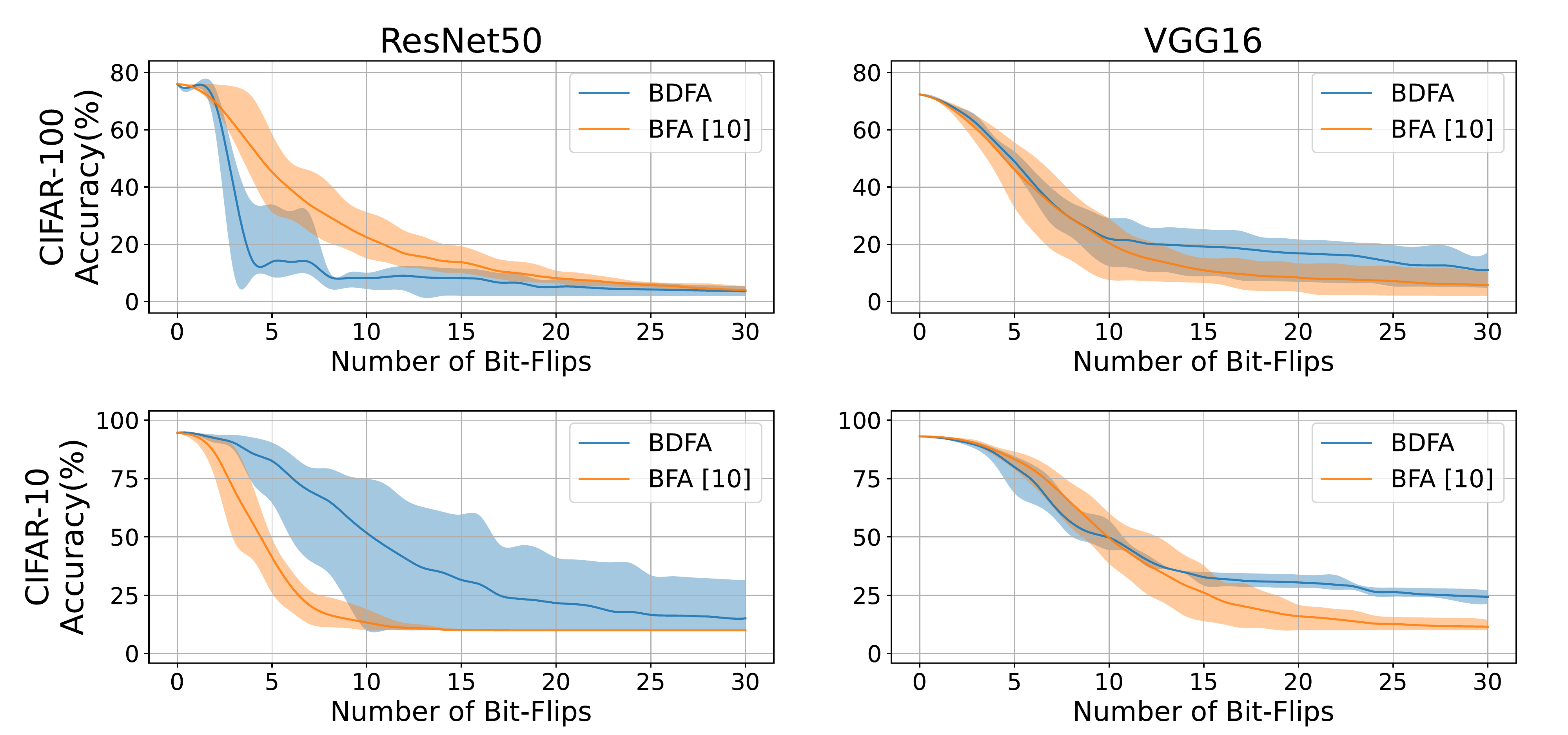}
 }
 
\caption{ The Top one accuracy of BDFA and BFA after performing 0 to 30 bit-flips. We repeated the experiment 5 times. The line indicates the average over these 5 tests. The shadow around each line indicates the error band(the minimum and maximum of these 5 tests).}

\label{figure:attackresults}
\end{figure*}

\subsection{Attack results and comparing to other methods}
In this section, we demonstrate our results on different networks and datasets and compare our results to previous work done by Rakin et al. \cite{rakin2019bit}. Table I shows the baseline accuracy of both networks on CIFAR-10 and CIFAR-100. In the experiments, we use a batch size of 128, the best batch size for Bit-Flip Attack \cite{rakin2019bit}, to have a fair comparison. However, we are able to generate data as much as we want.

\begin{center}
\label{T:baseline}
 \begin{tabular}{||c |c| c||} 
 \hline
 Network & CIFAR-10 Acc(\%) & CIFAR-100 Acc(\%) \\ [0.5ex] 
 \hline\hline
 Resnet50 & 94.63 & 75.96 \\ 
 \hline
 VGG16 & 93.05 & 72.34 \\ [0.5ex] 
 \hline
\end{tabular}
\captionof{table}{The baseline accuracy of 8 bit quantized Resnet50 and VGG16 on cifar10 and cifar100 datasets.}

\end{center}

\textbf{CIFAR100:} The top row of Figure \ref{figure:attackresults} shows the obtained results from attacking ResNet50 and VGG16 on the CIFAR-100 dataset. In ResNet50, the model accuracy decreases significantly from 75.96\% to 13.94\% with only 4 bits flips, and in 4 of the 5 tests performed, it reaches less than 9\%. Also, based on Table II, by continuing bit-flips up to 30 bits, it reaches an average of 3.6\%. For the VGG16 network, the model's accuracy after 9 bit-flips drops from 72.34\% to less than 30\% in all 5 tests, and the average accuracy in different runs after 30 bits-flips reaches 11.05\%, showing the attacks were quick and successful.


\begin{table}[!h]
\begin{center}
\begin{tabular}{|| c | c | c | c | c ||}
\hline
\textbf{Network} & \multicolumn{2}{ c |}{\textbf{CIFAR-100} }  & \multicolumn{2}{ c ||}{\textbf{CIFAR-10} }\\ 
\cline{2-5}
& \textbf{BDFA(\%)} & \textbf{BFA(\%)} & \textbf{BDFA(\%)} & \textbf{BFA(\%)} \\
\hline\hline
Resnet50  & 3.6 $\pm$ 1.6 & 3.94 $\pm$ 1.5 & 15.08 $\pm$ 16.3 & 10.1 $\pm$ 0.1 \\ \hline
VGG16     & 11.05 $\pm$ 6.3 & 5.8 $\pm$ 4.9  & 24.3 $\pm$ 2.9 & 11.5 $\pm$ 2.9\\ \hline
\end{tabular}
\captionof{table}{Top one accuracy comparison of BDFA and BFA after performing 30 bit-flips. These results are the average over 5 tests plus maximum error obtained by the average. .}
\vspace{-0.2in}
\end{center}
\label{T:equipos}
\end{table}

\textbf{CIFAR10:} The bottom row of Figure \ref{figure:attackresults} shows the accuracy drop comparison between BDFA and BFA \cite{rakin2019bit} on VGG16 and Resnet50 trained on CIFAR-10. As shown in Table II, the accuracy of VGG16 and ResNet50 decreases to nearly 24.3\% and 15.08\% with 30 bit-flips. This shows that BDFA is able to decrease the model accuracy significantly by only flipping 30 bits out of more than 500 million ResNet50 parameters(4 billion bits).

\textbf{Comparision to BFA:} As shown in Figure \ref{figure:attackresults}, both BDFA and BFA work well for finding the first few vulnerable bits and causing DNN to malfunction. In Resnet50 trained on CIFAR100, we achieve better performance than BFA, and with only 4 bit-flips, BDFA can decrease the model accuracy to 5-20\%. Therefore, it shows that the artificial data has better statistical similarities than one batch of training data. Although using distilled data can drastically decrease the accuracy of the model to 20-10\% with just 10-20 bit-flips, it can not completely destroy the function of DNN and decrease it to 0\%.


%% file: concl.tex
\section{Conclusion}
\label{sec:concl}

This paper presents a blind data bit-flip attack (DBFA) on deep neural networks, which exploits synthetic data for attack usage. We show that BDFA can decrease model accuracy dramatically to the random point. Experimental results show that the BDFA can perform similar results to data-dependent Bit-Flip Attack (BFA). We believe that we took the first step toward making a black-box adversarial parameter attack on deep neural networks.